\documentclass[11pt]{article}
\usepackage{amssymb}

\def\dd{\mbox{d}}

\def\o{\omega}

\def\l{\lambda}

\def\m{\mu}

\def\o{\omega}

\def\r{\rho}
\def\t{\tau}

\newcommand{\sm}[1]{\mbox{\scriptsize #1}}
\newcommand{\tn}[1]{\mbox{\tiny #1}}
\renewcommand{\@}[1]{\sqrt{#1}}

\renewcommand{\le}[1]{\label{#1}\end{eqnarray}}
\newcommand{\be}{\begin{equation}}
\newcommand{\ee}{\end{equation}}
\newcommand{\bea}{\begin{eqnarray}}
\newcommand{\eea}{\end{eqnarray}}
\newcommand{\nn}{\nonumber}

\newcommand{\eq}[1]{(\ref{#1})}
\def\nn{\nonumber\\}

\def\ffract#1#2{\raise .35 em\hbox{$\scriptstyle#1$}\kern-.25em/
\kern-.2em\lower .22 em \hbox{$\scriptstyle#2$}}

\def\half{{1\over2}\,}
\begin{document}

\begin{flushright}
AEI-2004-044\\
{\tt hep-th/0406093}\\
\end{flushright}
\vskip0.1truecm

\begin{center}
\vskip 2.5truecm {\Large \textbf{Brownian Motion, Chern-Simons Theory, and
2d Yang-Mills}}
\vskip 1.5truecm
%\vfill

{\large \textbf{Sebastian de Haro${}^{\star }${} and Miguel Tierz${}^{\dagger,\ddagger}$}}\\
\vskip .9truecm ${}^{\star }${\it Max-Planck-Institut f\"{u}r
Gravitationsphysik\\
Albert-Einstein-Institut, 14476 Golm, Germany}\\
\tt{sdh@aei.mpg.de} 
\vskip 4truemm 
$^\dagger${\it Applied Mathematics Department, The Open University\\
Walton Hall, Milton Keynes MK7 6AA, UK}\\
{\tt M.Tierz@open.ac.uk}
\vskip 4truemm
${}^{\ddagger }${\it Institut d'Estudis Espacials de Catalunya (IEEC/CSIC)\\
Edifici Nexus, Gran Capit\`{a}, 2-4, 08034 Barcelona, Spain}\\
{\tt tierz@ieec.fcr.es}
\end{center}

\vskip 2truecm

\begin{center}
\textbf{\large Abstract}
\end{center}

We point out a precise connection between Brownian motion, Chern-Simons
theory on $S^{3}$, and 2d Yang-Mills theory on the cylinder. The
probability of reunion for $N$ vicious walkers on a line gives the partition
function of Chern-Simons theory on $S^{3}$ with gauge group $U(N)$. The
probability of starting with an equal-spacing
condition and ending up with a generic configuration of movers gives the
expectation value of the unknot. The probability with arbitrary initial and
final states corresponds to the expectation value of the Hopf link. We find
that the matrix model calculation of the partition function is nothing but the additivity law of probabilities. We establish a correspondence between quantities in Brownian motion and the modular $S$- and $T$-matrices of the WZW model at finite $k$ and $N$. Brownian motion probabilitites in the affine chamber of a Lie group are shown to be related to the partition function of 2d Yang-Mills on the cylinder. Finally, the random-turns model of discrete random walks is related to Wilson's plaquette model of 2d QCD, and the latter provides an exact two-dimensional analog of the melting crystal corner. Brownian motion provides a useful unifying framework for understanding various low-dimensional gauge theories.

\newpage

\section{Introduction}

Low-dimensional gauge theories are a useful playground for understanding
quantum field theory. They also play an important role in string theory, and
in particular in topological string theories. 

Relations between statistical mechanical systems and quantum field theory are well-known and have proved useful on both sides (see, for example, \cite{brownianhepth}). A well-known example is provided by conformal field theories in two dimensions. For the case of higher-dimensional topological theories, however, such examples have until recently remained somewhat limited. An important recent breakthrough was the realization \cite{orv} that the closed topological A-model vertex discovered in \cite{topvertex} is related to a melting crystal corner \cite{inov}. In fact, D-branes can be included and have a natural interpretation as defects \cite{mc2}. In \cite{inov} this melting crystal picture was used as a definition of `quantum K\"ahler gravity', the quantum gravitational theory describing fluctuations of the K\"ahler structure and topology while keeping the complex structure fixed. In \cite{renate}, certain types of random walks were used to describe the combinatorics of triangulations of 2+1 quantum gravity. A review of the manifold connections between random walks/Brownian motion and conformal field theory and two-dimensional quantum gravity can be found in \cite{duplantier}.

In this letter we provide further examples of precise reformulations of
statistical mechanical systems in terms of gauge theories that are
topological or close to topological. Full details will be given elsewhere
\cite{dht}. Since the gauge theories in question have string
theory realizations \cite{wittstring,GoVa},
%(see \cite{marcos2} for a review)
it is our hope that this relation can be useful in string theory.
%\cite{adht}.

The first example relates Brownian motion of $N$ vicious --- that is,
non-intersecting --- walkers on a line
\cite{fisher} with certain boundary conditions -- or, equivalently, one
particle moving in the Weyl chamber of a simply-laced, compact Lie group $G$
-- to Chern-Simons theory \cite{witten} on $S^{3}$ for the corresponding
group. The correspondence works for the partition function, the expectation
value of the unknot, and the expectation value of the Hopf link invariant, the
basic reason being the correspondence between the WZW modular $S$- 
and $T$- matrices and Brownian motion quantities. The additivity law of 
probabilities can be reinterpreted as a matrix model derivation of the partition function of Chern-Simons theory on $S^3$, where the 
repulsive force of the Vandermonde interaction translates into an effective repulsion exerted by the walls of the Weyl chamber. The 
correspondence is at finite values of $k$ and $N$, where $N$ is the number of movers and $g_{s}={\frac{2\pi i}{k+N}}=-{\frac{1}{t}}$ is the inverse time. This is reminiscent of relation found in \cite{orv}, where $t$
played the role of the temperature.

The second example concerns a walker moving in the fundamental Weyl chamber
of an affine Lie algebra. This can be reinterpreted as the partition 
function of 2d Yang-Mills on the cylinder, where the initial and final positions
correspond to states at the two ends of the cylinder, and now time is 
related to the area. Our derivation reproduces formulas in the mathematical literature which make the structure of the partition function much more transparent than the usual sum over representations. In particular, the partition function is shown to be an affine character, and so the modular properties of 2d YM on the cylinder are most explicit in this formulation.

The third example is the case of discrete random walkers rather than
continuous Brownian motion; we analyze the random turns model \cite
{fisher,forrester}. Its relation to randomly growing Young tableaux and (by
Poissonization)
%\footnote{%
%Poissonization is essentially a way to map recursive combinatorial 
%sequences
%into generating functions (Poisson transform). Detailed information about
%Poissonization and de-Poissonization can be found in \cite{Flaj}, for
%example.})
to lattice QCD$_{2}$ are known in the mathematical literature. We reinterpret the lattice QCD$_{2}$ as a two-dimensional melting crystal,
where the energy cost for removing a particle is the (logarithm of) the gauge
coupling.

\section{Brownian motion and Chern-Simons theory}\label{cs}

In this paper we will be concerned with Brownian motion of $N$
movers on a line. We will study so-called \textit{vicious walkers} \cite
{fisher}. These are random walkers whose trajectories are not allowed to
intersect. As is well-known, the intersection properties of such a random
walk process play an important role in quantum field theory (as reviewed in
detail in \cite{duplantier,brownianhepth}). 
%Note that the
%difference between Fisher's vicious random walkers and the usual
%non-intersecting random walks, that play a crucial role in connection with
%the $O(N)$ model and its applications to polymer physics \cite
%{duplantier,brownianhepth}, is that the former are \emph{directed}
%non-intersecting random walks (for details, see second reference in 
%\cite{duplantier}).

Let us first consider the case of a single free walker performing Brownian
motion on a line. The probability for it to travel from $y$ to $x$ in time 
$t$ is given by
\begin{equation}
p_{t}(x,y)={\frac{1}{\sqrt{4\pi Dt}}}\,e^{-(x-y)^{2}/4Dt}~.
\end{equation}
Here, $D$ is the diffusion coefficient of the medium, which from now on we
will set equal to $\half$. This probability, and its
generalizations, is the basic quantity that we will study in this paper. It
can be obtained by a variety of methods; for an overview see e.~g.~\cite
{brownianhepth,brownianstat}. Let us however mention some of its important
properties. First of all, it can be obtained as the continuum limit of a discrete random walk, where at each tick of the clock the particle can travel a fixed finite distance left or right with equal probability (see also section~\ref{qcd2}). The probability is then a binomial distribution whose continuum limit is Gaussian. Further, $p_{t}(x,y)$ satisfies the diffusion or heat equation
\begin{equation}
{\frac{\partial }{\partial t}}\,p_{t}(x,y)=D\Delta p_{t}(x,y),
\end{equation}
where $\Delta $ is the Laplacian in $x$. Finally, $p_{t}(x,y)$ turns into a 
Dirac delta function of the position $x-y$ when $t$ tends to zero.

The above has an obvious higher-dimensional generalization:
\begin{equation}
p_{t,N}(x,y)={\frac{1}{(2\pi t)^{N/2}}}\,e^{-{\frac{|x-y|^{2}}{2t}}},
\end{equation}
and $N$ is the dimension. Equivalently, this can be regarded as the
product of the probabilities of $N$ single movers on a line, i.e.~as the
probability for $N$ movers on a line to start at positions $y_{1},\ldots
,y_{N}$ and end up at $x_{1},\ldots ,x_{N}$ after time $t$.

The particular process that we will relate to Chern-Simons theory is that of
$N$ vicious walkers on a line. Walkers are vicious \cite{fisher} if they
annihilate each other when they meet. Thus, we will impose a
non-intersecting condition and compute a probability for these walkers to
walk from one configuration to another without ever intersecting. If we
denote their coordinates by $\lambda_i$, $i=1,\ldots,N$, they satisfy 
$\lambda_1>\lambda_2>\ldots>\lambda_N$. Alternatively, this process can be
regarded as motion of a single particle in the fundamental Weyl chamber of 
$U(N)$. The particle starts moving at position $\mu_i$ satisfying $
\mu_1>\mu_2>\ldots>\mu_N$, and is required to stay within the Weyl chamber.
The process stops when the particle hits one of the walls. One then computes
the probability of going from an initial position $\mu_i$ to a final
position $\lambda_i$ staying always within the chamber. This is given by
\cite{fisher}:
\begin{equation}
p_{t,N}(\lambda,\mu)={\frac{1}{(2\pi t)^{N/2}}}\,e^{-{\frac{%
|\lambda|^2+|\mu|^2}{2t}}} \,\det|e^{\lambda_i\mu_j/t}|_{1\leq i<j\leq N}~.
\end{equation}
Obviously, this probability is symmetric under interchange of initial and
final boundary conditions.

Let us now evaluate this amplitude in a specific case. We take the same initial and final boundary conditions, i.e. $\m=\l$, and further an equal spacing condition, that is, $\l_{0j}-\l_{0,j+1}=a$, where $a$ is the initial and final spacing between two neighboring movers. We compute the so-called probability of reunion --- the probability that the movers go back to their (almost coinciding) positions after time $t$. Of course, this is an exponentially vanishing probability. Notice that, since the $\l$'s also label highest weights of irreducible representations of $U(N)$, this boundary condition is labeled by the Weyl vector for a suitable choice of the overall scale. Now a straightforward computation yields
\begin{equation}\label{prob00}
p_{t,N}(\lambda _{0},\lambda _{0})={\frac{1}{(2\pi t)^{N/2}}}\,
\prod_{k=1}^N(1-e^{-ka^2/t})^{N-k}~.
\end{equation}
We are going to relate this probability to the partition function of 
Chern-Simons theory. 

Recall that Chern-Simons theory is 
a topological quantum field theory whose action is built of a Chern-Simons 
term involving as gauge field a gauge connection associated to a group $G$ on 
a three-manifold $M$ \cite{witten} (see \cite{marcos2} for a recent review). 
The action is:
\begin{equation}
S(A)={k \over 4 \pi} \int_M {\rm Tr} \Bigl( A \wedge \dd A + {2 \over 3}\, A
\wedge A \wedge A \Bigr)~,   
\end{equation}
with $k$ an integer number. Now if we choose units where $a^2=1$ and identify
\begin{equation}
-{\frac{1}{t}}=g_s={\frac{2\pi i}{k+N}}~,
\end{equation}
with $g_s$ the string coupling, equation \eq{prob00} is the partition 
function of Chern-Simons on $S^{3}$ with gauge group $U(N)$ \cite{witten}.
Observables in Chern-Simons theory always come with a choice of framing, 
corresponding to a choice of trivialization of the tangent bundle in the
gravitational Chern-Simons term \cite{witten}. In our case, notice that the 
framing is the matrix model framing \cite{akmv,miguel}, which is related to the
canonical framing as follows:
\begin{equation}
Z_{\mbox{\scriptsize CS}}(S^{3})=e^{{\pi i\over2}\,N^2}q^{-{1\over12}\,
N(N^2-1)}\,p_{t,N}(\l_0,\l_0)~,
\end{equation}
where the label $0$ refers to the Weyl vector $\l_0=\r$, and as usual 
$q=e^{g_s}$.

One way to understand why we get the partition function of
Chern-Simons theory is to generalize the above to other compact groups \cite
{fisher,grabiner}. Using the method of images, one finds that the above
probability generalizes to
\begin{equation}
p_{t,r}(\l,\m)={\frac{1}{(2\pi t)^{r/2}}}\,e^{-{\frac{|\lambda |^{2}+|\mu 
|^{2}}{2t}%
}}\sum_{w\in W}\epsilon (w)e^{(\lambda ,w\mu )/t},
\end{equation}
where $r$ is the rank of $G$ and $W$ the Weyl group. From here,
using the Weyl denominator formula we can get amplitudes for more general
boundary conditions:
\be\label{unknot}
p_{t,r}(\lambda ,\rho )={\frac{1}{(2\pi t)^{r/2}}}\,
e^{-{\frac{|\lambda|^{2}+|\rho |^{2}}{2t}}}
\prod_{\alpha >0}2\sinh {\frac{(\alpha ,\lambda)}{2t}}~,
\ee
where $\alpha $ are the positive roots and $\rho $ is the Weyl vector. This
expression is the (unnormalized) expectation value of a Wilson loop around the
unknot. The partition function is obtained by setting $\l=\r$. Normalizing 
\eq{unknot} by the partition function gives the quantum dimension.

Somewhat more generally, taking into account the framing and the central
charge we can in fact write, dropping an overall sign,
\begin{equation}
p_{t,r}(\l,\m)=e^{{2\pi i\over12}\,{\sm{dim}}\,g}\,
(TST)_{\lambda \mu },
\end{equation}
where
\begin{eqnarray}
S_{\lambda \mu } &=&{\frac{i^{|\Delta _{+}|}}{(k+g)^{r/2}}}\,|P/Q^{\vee 
}|^{-{\frac{1}{2}}\,}
\sum_{w\in W}\epsilon (w)e^{-{2\pi i\over k+g}(\lambda ,w\cdot \mu )}
\nonumber \\
T_{\lambda \mu } &=&\delta _{\lambda \mu }\,e^{{\frac{2\pi iC(\lambda )}{%
2(k+g)}}-{\frac{2\pi ic}{24}}}~.
\end{eqnarray}
The central charge is $c=k\,{\mbox{dim}}\,g/(k+g)$, $C(\lambda )$ is the Casimir
of the representation $\lambda $, $\Delta _{+}$ is the set of positive
roots, $P$ is the weight lattice, and $Q^{\vee }$ is the coroot lattice.
Obviously, $S$ is the Brownian motion probability (with the external factors
of $T$ amputated), and $T$ is the Boltzmann factor. It is now also clear 
that
$p(\l,\m)$ itself corresponds to the (unnormalized) expectation value of the
Hopf link invariant with representations $\l$ and $\m$. $TST$ is in fact the 
operator that performs the modular transformation $\t\rightarrow\t/(\t+1)$
(see also \cite{akmv}). By this particular surgery \cite{witten} one obtains 
$S^3$ out of two solid tori.

To end this section, let us remark that there is a matrix model expression
for the partition function of Chern-Simons on $S^{3}$ \cite{marcos,miguel}:
\begin{equation}
Z_{\mbox{\scriptsize CS}}(S^{3})={\frac{e^{-{1\over12}\,N(N^2-1)\,g_s}}{N!}}\,
\int \prod_{i=1}^{N}{\frac{\mbox{d}\lambda _{i}}{2\pi }}
\,e^{-|\lambda |^{2}/2g_{s}}\prod_{i<j}\left( 2\sinh {\frac{\lambda
_{i}-\lambda _{j}}{2}}\right) ^{2}~.
\end{equation}
It is not hard to check that this is nothing but the extensivity property of
probabilities:
\begin{equation}
p_{t+t^{\prime },r}(\rho ,\rho )=\int [\mbox{d}\lambda ]\,p_{t,r}(\rho
,\lambda )\,p_{t^{\prime },r}(\lambda ,\rho ),  \label{additivity}
\end{equation}
where the range of integration is the same as in the matrix model, but using
symmetry can be restricted to the Weyl chamber. This can also be seen as a
renormalization group property. It is also clear that the repulsive $\sinh$, coming from the Vandermonde determinant, is related to 
the fact that the walls of the Weyl chamber are effectively {\it repeling}. Indeed, all the paths that end on these walls are suppressed from the expression for the probability, and hence only paths will contribute for which the particle stays away from the walls.

We saw that the partition function comes out in the natural matrix model
framing. It would be interesting to see if one can obtain more generic
framings by rescalings and shifts of the boundary conditions. For the
case of the partition function and the unknot this seems possible 
\cite{dht}. This results in real exponential factors, as expected from the analytic continuation of the phase factors.

\section{Brownian motion and QCD$_2$}\label{qcd2}

It is now natural to ask what happens if we consider Brownian motion in the fundamental Weyl chamber of an affine Lie algebra. The affine Weyl group is $\tilde{W}=W\ltimes T$, where $T$ is the group of translations in root space. In the case analyzed in the previous section, constraining the motion to the fundamental Weyl chamber was achieved by appropriately adding all images generated by the action of $W$. Now, in addition, we have to mod out by translations in the coroot lattice. We get the following density
\begin{equation}
q_{t,r}(\lambda ,\mu )={\frac{1}{(2\pi t)^{r/2}}}\sum_{\gamma \in lQ^{\vee
}}\sum_{w\in W}\epsilon (w)e^{-{\frac{1}{2t}}|\gamma +\lambda -w\cdot \mu
|^{2}},
\end{equation}
where $r$ is the rank and $l=k+g$. Standard manipulations yield
\begin{equation}\label{qtr}
q_{t,r}(\lambda ,\mu )={\frac{1}{(2\pi t)^{r/2}}}\sum_{w\in \tilde{W}}e^{-{%
\frac{1}{2t}}|\hat{\lambda}-w\cdot \hat{\mu}|^{2}},
\end{equation}
for affine vectors $\hat{\lambda}=\lambda +l\hat\o_0$, following common
notation $\hat\o_0=(0;1;0)$. In fact, the
normalized density is the affine character ${\mbox{ch}}_{\hat{\l}}
(\hat{\mu}/t)$, in complete analogy with the finite case.
With our boundary conditions, formula \eq{qtr} agrees with the general result 
found in \cite{gesselzeilberger}.

Using the results in \cite{frenkel}, it is not hard to see that the above
is related to the partition function of 2d Yang-Mills theory on
the cylinder. Two-dimensional Yang-Mills (see \cite{moore} for a review) is
not a topological theory, but it is close to that in the sense that it has 
no
local degrees of freedom. It is invariant under area-preserving
diffeomorphisms. On the cylinder, the partition function is \cite{migdal}:
\begin{equation}
Z_{\mbox{\scriptsize 2dYM}}(g,g^{\prime };t)=\sum_{\l\in P_{++}}
\chi_\l(g^{-1})\chi_\l(g^{\prime })\,e^{-{\frac{t}{2}}\,C(\lambda )
-{t\over2}\,|\r|^2}~,
\end{equation}
where $P_{++}$ are the dominant weights\footnote{For convenience we included a 
constant factor of $|\r|^2$.}. The gauge coupling $g_{\mbox{\tn{YM}}}$ 
and the area of the cylinder $A$ always appear through the combination 
$t=g_{\mbox{\tn{YM}}}^{2}A$.

From the cylinder one can easily obtain the partition function for other 
two-dimensional manifolds. A salient feature of the partition function on the 
cylinder is its
extensivity. If we glue two cylinders of areas $A_1$ and $A_2$ along a
common boundary, the partition function on the resulting cylinder of area $%
A^{\prime}=A_1+A_2$ will have the same form, with new area $A^{\prime}$.
An analogous property also holds on the disk, and can be seen as an 
invariance under renormalization group transformations and indicates that 
the plaquette model for this theory is in fact exact. Notice that the area 
is linearly related
to the Brownian motion time $t$, and so this is like the additivity property
(\ref{additivity}) that we encountered before.

As is well-known, the partition function of 2d Yang-Mills is a solution of
the heat equation on a group manifold. It is the kernel of the heat equation
defined such that it satisfies the heat equation for both $g$ and 
$g^{\prime}$,
and in the limit $t\rightarrow0$ it tends to a Dirac delta function $\delta(g-g^{\prime})$ with respect to the Haar measure. From it one can construct the unique solution of the heat equation with specified boundary condition at $t=0$ \cite{fegan1}.

Let us now state the precise relation between Brownian motion and 2d
Yang-Mills. We consider the group elements $g=e^{2\pi i\l/l}$,
$g^{\prime}=e^{2\pi i\m/l}$. Using
the manipulations in \cite{frenkel}, we find
\begin{equation}
Z_{\mbox{\scriptsize 2dYM}}(g,g';t)={\frac{(-il)^r\,\mbox{vol}\,(P/Q^\vee)}
{D_\r(2\pi i\l/l)D_\r(-2\pi i\m/l)}}\,\,
q_{t,r}(\lambda,\mu)~,
\end{equation}
where $q_{t,r}$ is given in \eq{qtr}. The normalization factors $D_\r$ come from the normalization of the characters that enter $Z_{\mbox{\scriptsize 2dYM}}$,
whereas $q_{t,l}$ is always unnormalized, as we also saw in the previous
section. In particular, the normalization is independent of $t$ and thus the $t$-dependence is totally contained in $q_{t,r}$. The above relation is our main result in this section.

It is well-known that the partition function of 2d YM on the sphere does not
have nice modular transformation properties. The reason is that it is
proportional to the derivative of a modular form, rather than the modular
form itself. For the case of the torus, see \cite{vafa,rudd}. The modular properties of the partition function on the cylinder are easy to work out from the above. It suffices to realize that $q_{t,r}(\l,\m)$ is an affine character, which is obtained from a theta function by summing over all images. We hope to come back to this issue \cite{dht}. Related work appeared in \cite{zelditch}.
Let us here remark that there is a special group element which one can insert, so that the partition function on the sphere does have nice modular properties. This is the special group element defined in \cite{kostant,fegan1} such that the character has values $0,\pm1$. Let us denote by $|a\rangle$ the associated state. We can then interpret 
$Z_{\mbox{\scriptsize 2dYM}}(a;t)$ as a partition function on the sphere 
with insertion of a state $|a\rangle$ at $t=0$. The partition function can 
be written in terms of the Dedekind $\eta$-function by Macdonald's $\eta$-function formula 
\cite{macdonald}:
\begin{equation}
Z_{\mbox{\scriptsize 2dYM}}(a;t)=e^{-{\frac{t}{24}} \,{\sm{dim}\,G}} 
\,\eta(t)^{\sm{dim}\,G}~.
\end{equation}

We will end this section with some comments on one-dimensional discrete random 
walks and their relation to growing Young tableaux and Wilson's plaquette
model\footnote{There are various possible reformulations of the discrete
random walks problem, that we will not consider here: as a Hammersley process,
a growing PNG droplet, etc. See e.g. \cite{randomwalks}.}. 
Brownian motion is a limiting case of this \cite{fisher}. We will focus on the random-turns model \cite{fisher,forrester}, where $N$ movers are allowed to move on a (one-dimensional) lattice, but at each tick of the clock only one mover 
walks, and it can take a step left or right with equal probability. 
It is known \cite{rains,forrester} that the probability of reunion after $2n$
steps in this model is given by the expectation value of $2n$ powers of a 
unitary matrix in a unitary ensemble of $N\times N$ matrices. This probability
is also proportional to the probability of the largest increasing subsequence 
of a random permutation of $n$ objects of having length $\leq N$. By the
Schensted correspondence between random permutations and Young diagrams,
this is also the probability for the top row of a Young diagram with $n$ boxes
to have length $\leq N$. Let us call this probability $P(l_n\leq N)$. The 
Gross-Witten model is obtained by studying the Poissonized quantity, i.e.
studying all possible (Poisson distributed) Young tableaux whose top row has
length $L_\l\leq N$:
\bea
P(L_\l\leq N)&=&e^{-\l}\sum_{n=0}^\infty{1\over n!}\,\l^n\,P(l_n\leq N)\nn
&=&e^{-\l}\sum_{n=0}^\infty \sum_{\m\vdash n\!,\,\m_1\leq N}{1\over n!}\,\l^n{d_\m^2\over n!},
\eea
where $L_\l$ is the length of the top row of a set of diagrams that are
Poisson distributed, and $\m\vdash n$ means that $\m$ partitions $n$, and 
$d_\m$ is given by the hook formula of the diagram. But 
$d_\m^2\over n!$ is the Plancherel measure, which is also the probability to 
pick a Young diagram of shape $\m$ among a random set. Thus, the above is the
grand canonical ensemble distribution for tableaux with top rows of lengths
$L_\l\leq N$, and $\log\l$ is the chemical potential for adding a box anywhere
in the diagram so that it remains a valid $U(N)$ tableau. If we identify $\l$
with the gauge coupling, we obtain Wilson's lattice version of QCD$_2$
\begin{equation}\label{GW}
Z_{\mbox{\scriptsize GW}}=\int \mbox{d}U\exp [{\frac{1}{g^{2}}}
{\mbox{Tr}}_{\sm{F}}(U+U^{\dagger })]=e^\l P(L_\l\leq N) =\sum_{n=0}^{\infty } 
{\frac{1}{(2n)!}} {\frac{1}{g^{4n}}}\,Z_{2n}(\mu _{j}=j,\lambda _{i}=i)~,  
\end{equation}
where $\l=1/g^4$. The trace is taken in the fundamental representation of 
$U(N)$. The last equality relates the partition function to the random walks probability. The plaquette model was solved at finite $N$ in \cite{barsgreen}. In \cite{gw} it was found that the model has a third-order phase transition\footnote{It is not hard to see that the phase transition precisely comes from the restriction on the number of boxes in the top row to be $\leq N$.} at large $N$ . Thus, lattice QCD$_2$ can be reformulated as a growing (or shrinking) Young tableau.

The lattice model of QCD$_2$ is celebrated for its phase transition. This
occurs at $\lambda=g^2N=2$ and is closely related to a similar phase
transition in the probabilities of the random distribution. Indeed, \cite
{johansson1} has proved a depoissonization lemma that bounds the value of
the probability of the random distribution from above and from below with
its Poissonized value. The phase transition for the random Young diagram
occurs when the number of boxes grows as $n\sim2\sqrt{N}$.

Finally, let us comment that also in the case of the growing Young tableau
there is a limiting shape after appropriate rescaling with $N$. This
limiting shape is given by a continuous (but
non-differentiable) function, the discontinuity being of course at the point
of the phase transition. This is a two-dimensional analogue of the limiting
shape of the 3d partitions found in \cite{orv}.

\section{Discussion and outlook}

In this letter we have pointed out three connections: between Brownian
motion in the fundamental Weyl chamber of a simply-laced, compact Lie group,
and Chern-Simons theory on $S^{3}$ for the corresponding group; between
Brownian motion in the Weyl chamber of an affine Lie group and 2d Yang-Mills
theory; and between the random-turns model of discrete random walks, a
two-dimensional melting crystal, and lattice QCD$_{2}$. The fact that the
connections are quite general --- in the case of Chern-Simons, we get a
statistical mechanical realization of the modular matrices --- and work for
various gauge groups, might lead one to think that there may be more to the
relation between random/diffusion walks and gauge theories
than just a mathematical coincidence, and one might hope to find physical
applications. Therefore these connections immediately raise several
questions. First of all, how far does the correspondence go? In the case of
Chern-Simons, we saw that the agreement can be understood from the
representation of the modular $S$- and $T$-matrices as non-intersecting
Brownian motion probabilities and Boltzmann factors, respectively. Taking
into account the role of $SL(2,{\mathbb Z})$ in the surgery approach of
Chern-Simons theory \cite{wittstring}, one may hope that for manifolds other than $S^3$ at least the simplest cases may have an interpretation in terms of Brownian motion quantities. This is certainly worth exploring further.
%Preliminary computations shows that this seems to be the case
%for lens spaces, while more complicated rational homology spheres (e.g. 
%like
%the Poincare homology sphere) may be in jeopardy. However, in this sense,
%recall that in general one does not expect the Chern-Simons/topological
%string theory correspondence to hold for any of these above-mentioned
%manifolds. In addition, it has been recently found \cite{LawZag} that
%already the partition function looses part of its modular properties when
%the Seifert manifold is more complicated than a lens space.
Also, the basic cases of the partition function, the unknot and the Hopf link can be easily obtained from Brownian motion. It would be interesting to see if more general knots can also be obtained.

In the affine case, where one finds the partition function of 2d Yang-Mills
on the cylinder, it would be interesting to see if one can extend the
connection to expectation values of Wilson lines, or whether
there is a Brownian motion interpretation of the partition function on the
three-punctured sphere.

A particularly interesting point would be to see if Brownian motion can give
us further connections between these low-dimensional theories. For example,
the fact that Brownian motion is a limit of a discrete random walk is very
suggestive of a connection between two-dimensional and three-dimensional
theories. Also, it would be interesting to see whether the lock-step model
\cite{fisher}, which we have not considered in this paper, also has a
reformulation in terms of a low-dimensional gauge theory. Notice also that the way 2d Yang-Mills arises from Brownian motion is by effectively compactifying the dual Cartan subalgebra to a torus. The resulting expression is an affine character, which makes the modular properties of the partition function completely explicit. A connection between 2d Yang-Mills on the torus and topological strings has recently been pointed out in \cite{vafa}.

We saw that the matrix model of Chern-Simons theory on $S^{3}$ naturally
arises in the composition law of probabilities. Indeed, since we are dealing
with continuous paths it seems that the matrix model formulation of
Chern-Simons theory is the most natural one for Brownian motion. Yet from the point of view of the WZW model one naturally gets sums over representations rather than integrals. At the level of the intermediate states, there is a precise way in which both approaches are equivalent \cite{dht}. In Chern-Simons theory the representations are integrable, and from the Brownian motion point of view these correspond to special points on the line. We can deal more generally with arbitrary points by using characters. It would also be interesting to explore the connection with the fermionic representation.

On the more mathematical side, the underlying principle allowing these
connections seems to be the fact that all these models in one way or another
satisfy the heat equation. It has been known for a long time that certain
quantities in the WZW model satisfy the heat equation \cite{bernard}. Also,
the heat equation is closely related to modular invariance. We hope to
report more on this in the future \cite{dht}.

Perhaps one of the most interesting questions is whether Brownian motion can
be used as a tool in string theory, in the spirit of \cite{orv,inov}, for
example. Chern-Simons theory is the effective gauge theory describing the
topological A-model \cite{wittstring}, and so a reformulation in terms of
Brownian motion might be useful for string theory itself. Notice furthermore
that the natural string theory coupling is related to the Brownian motion
parameter $t$ (which is actually the product of the time parameter and the
diffusion coefficient) as $-{\frac{1}{t}}=g_{s}$, whereas the relation to the
Chern-Simons coupling involves analytic continuation\footnote{%
This analytic continuation is a subtle issue
%-already present in many works
%dealing with Chern-Simons or with topologically 
%massive field theories-,
that we hope to get back to in the future.}. This suggests that the
interpretation of topological strings in terms of a statistical mechanical
system may in some ways be more natural than as a gauge theory. In
particular, it would be extremely interesting to understand whether the heat
equation plays a role in the topological A-model. One would also like to see
if discrete random walks --- of which Brownian motion is a limit --- are
related to topological strings. Notice that \cite{nekrasov} have used
random partitions --- which, as pointed out, are equivalent to the random
walks model and, after Poissonization, to the plaquette model of QCD$_{2}$
--- to compute the prepotential of $\mathcal{N}=2$ SYM theory.

Another interesting question is whether 2d Yang-Mills and lattice QCD$_2$,
and their respective third-order phase transitions, have string theory
interpretations. In \cite{dv2} it was shown that the QCD$_2$ plaquette model
can be used to obtain the $SU(2)$ $\mathcal{N}=2$ Seiberg-Witten solution by
taking a double scaling limit. In particular, a local Calabi-Yau geometry
that engineers this curve was found. In this case, it was argued that the
phase transition plays no role. In \cite{dvdeconstruction}, 2d Yang-Mills on
a Riemann surface $\Sigma$ was obtained by wrapping D6 branes on $
S^1\times\Sigma$.

We saw that the Gross-Witten phase transition does play a role in the
context of the shrinking two-dimensional Young tableau. It was related to 
the
non-differentiability of the limiting shape. It would be interesting to see
whether such phase transitions are present and play any role for
the topological vertex.

Let us also mention that Cardy \cite{Cardy1} has recently found
a remarkable connection between a system of $N$ non-intersecting Brownian
motions (described through the celebrated SLE process) and boundary-bulk
conformal field theory models and integrable models of Sutherland type. 
%To see whether there is a connection 
%with our work seems also worth exploring.

Finally, recall that \cite{witten1} pointed out connections between
vertex models and Chern-Simons theory. The random walks that we have
mentioned in this paper are special cases of vertex models. However, for us
the connection with Chern-Simons theory appears in the continuous limit of
Brownian motion rather than in the discrete case.

\section*{Acknowledgments}

We thank Mina Aganagic, Bernard de Wit, Robbert Dijkgraaf, Kirill Krasnov, Renate Loll, Marcos Mari\~{n}o, Matthias Staudacher and Stefan Theisen for interesting discussions and comments on the paper. We also thank each other's institutes for hospitality at various stages of this work.

\end{document}